\documentclass[10pt,a4paper]{article}

\usepackage[utf8]{inputenc}
\usepackage[english]{babel}
\usepackage{amssymb,amsmath}
\usepackage{euscript}

\begin{document}
\large

\newpage
\begin{center}
\LARGE{\bf An Axial-Vector Photon in a Mirror World}
\end{center}
\vspace{0.1mm}
\begin{center}
{\bf Rasulkhozha S. Sharafiddinov}
\end{center}
\vspace{0.1mm}
\begin{center}
{\bf Institute of Nuclear Physics, Uzbekistan Academy of Sciences,
\\Ulugbek, Tashkent 100214, Uzbekistan}
\end{center}
\vspace{0.1mm}

\begin{center}
{\bf Abstract}
\end{center}

The unity of symmetry laws emphasizes, in the case of a mirror CP-even Dirac Lagrangian, the regularity that the left- and right-handed axial-vector photons refer to long- and short-lived 
bosons of true neutrality, respectively. Such a difference in lifetimes expresses the unidenticality of masses, energies, and momenta of axial-vector photons of the different components. They require the generalization of the classical Klein-Gordon equation to the case of C-odd types of particles with a nonzero spin. Together with a new Dirac equation for truly neutral particles with the 
half-integral spin, the latter reflects the availability in nature of the second type of the 
local axial-vector gauge transformation responsible for origination in a Lagrangian of C-oddity of 
an interaction Newton component, which gives an axial-vector mass to all the interacting particles and fields. The quantum axial-vector mass, energy, and momentum operators constitute herewith the 
CP-invariant Schr\"odinger equation, confirming that each of them can individually influence on the matter field. Thereby, findings define at the new level, namely, at the level of the mass-charge structure of gauge invariance the mirror Euler-Lagrange equation such that it has an axial-vector nature. 

\vspace{0.3cm}
\noindent
{\bf Key words:} Mass, Energy, and Momentum Matrices; C-Odd Particles with a Nonzero Spin; Mass, Energy, and Momentum Operators; Gauge Transformations of the Second Type; An Interaction Newton Component; Selected Quantum Theory Equations. 

\vspace{0.3cm}
\noindent
{\bf PACS:} 03.65.Pm, 11.15.-q, 11.30.Hv, 11.40.-q,	04.20.Gz

\vspace{0.8cm}
\noindent
{\bf 1. Introduction}
\vspace{0.4cm}

Between nature of elementary particles and matter fields there exists a range of fundamental symmetries, which require to raise the question about axial-vector photons having with truly 
neutral fermions a C-noninvariant interaction. Their presence [1] restores herewith the broken 
gauge invariance of the unified field theory Dirac Lagrangian of C-odd particles.

As a consequence, the left (right)-handed neutrino of true neutrality in the field of an 
axial-vector emission can be converted into a right (left)-handed one without change of his own flavor. These interconversions together with the unity of flavor and gauge symmetry laws [1,2] express the unidenticality of masses, energies, and momenta of truly neutral neutrinos of the  different components. However, such a possibility, as was noted in [3] for the fist time, is realized  
only at the spontaneous mirror symmetry violation of axial-vector types of fermions. In other words, the left-handed neutrino of true neutrality and the right-handed axial-vector antineutrino are of long-lived leptons of C-oddity, and the right-handed truly neutral neutrino and the left-handed axial-vector antineutrino refer to short-lived C-odd fermions.

This difference in lifetimes establishes a new CP-even Dirac equation and thereby describes 
a situation when the mass, energy, and momentum come forward in nature of truly neutral types 
of particles as the flavor symmetrical matrices
\begin{equation}
m_{s}={{0 \ \, \, \, \, m_{A}}\choose{m_{A} \, \, \, \, \ 0}}, \, \, \, \,
E_{s}={{0 \ \, \, \, \, E_{A}}\choose{E_{A} \, \, \, \, \ 0}}, \, \, \, \,
{\bf p}_{s}={{0 \ \, \, \, \, {\bf p}_{A}}\choose{{\bf p}_{A} \, \, \, \, \ 0}},
\label{1}
\end{equation}
\begin{equation}
m_{A}={{m_{L} \, \, \, \, 0}\choose{\ 0 \, \, \, \, \ m_{R}}}, \, \, \, \,
E_{A}={{E_{L} \, \, \, \, 0}\choose{\ 0 \, \, \, \, \ E_{R}}}, \, \, \, \,
{\bf p}_{A}={{{\bf p}_{L} \, \, \, \, 0}\choose{\ 0 \, \, \, \, \ {\bf p}_{R}}},
\label{2}
\end{equation}
where an index $A$ denotes the block matrix.

But here we must recognize that the same particle has no simultaneously both vector C-even and 
axial-vector C-odd charges. Such an order, however, corresponds to the fact that the same photon 
may not be simultaneously both a vector gauge boson and an axial-vector one. Thereby, it opens in principle the possibility for the classification of elementary objects with respect to C-operation, which reflects the availability [1,2] of the two types of particles and fields of C-invariant and 
C-noninvariant nature. 

The mass, energy, and momentum of the neutrino of a C-even charge are strictly vector $(V)$ type [4]. In contrast to this, the neutrino of a C-odd electric charge [5] has the mass, energy, and momentum 
of an axial-vector $(A)$ nature [3]. Therefore, the matrices (\ref{1}) and (\ref{2}) undoubtedly refer only to those elementary particles in which the vector C-even properties are absent. 

Of course, these matrices from the point of view of nature itself give the right to write the 
unified field theory equation of truly neutral types of particles with the spin $1/2$ as a unification of the structural parts of their four-component wave function 
$\psi_{s}(t_{s}, {\bf x}_{s})$ in a unified whole
\begin{equation}
i\frac{\partial}{\partial t_{s}}\psi_{s}=\hat H_{s}\psi_{s}.
\label{3}  
\end{equation} 

So it is seen that 
\begin{equation}
\hat H_{s}=\alpha \cdot\hat {\bf p_{s}}+\beta m_{s},
\label{4}
\end{equation}
and the sizes of $m_{s},$ $E_{s}$ and ${\bf p}_{s}$ correspond in a mirror presentation [3] of 
matter fields to the quantum axial-vector mass, energy, and momentum operators 
\begin{equation}
m_{s}=-i\frac{\partial}{\partial \tau_{s}}, \, \, \, \,  
E_{s}=i\frac{\partial}{\partial t_{s}}, \, \, \, \,  
{\bf p}_{s}=-i\frac{\partial}{\partial {\bf x}_{s}}.
\label{5}
\end{equation}

The presence of an index $s$ in (\ref{1}), (\ref{4}), and (\ref{5}) implies the unidenticality 
of the space-time coordinates $(t_{s}, {\bf x}_{s})$ and the lifetimes $\tau_{s}$ for the left 
$(s=L=-1)$- and right $(s=R=+1)$-handed particles. Then it is possible, for example, to use [3] 
any of earlier experiments [6,7] about a quasielastic axial-vector mass as the first laboratory indication in favor of an axial-vector mirror Minkowski space-time.

We see in addition that the Dirac matrices $\gamma^{\mu}=(\beta, \beta \alpha)$ for the case 
$\partial_{\mu}^{s}=\partial/\partial{\it x}^{\mu}_{s}=(\partial/\partial t_{s}, -\nabla_{s})$ when 
\begin{equation}
\partial_{\mu}^{s}=
{{0 \ \, \, \, \, \partial_{\mu}^{A}}\choose{\partial_{\mu}^{A} \, \, \, \, \ 0}}, \, \, \, \,
\partial_{\mu}^{A}=
{{\partial_{\mu}^{L} \, \, \, \, 0}\choose{\ 0 \, \, \, \, \ \partial_{\mu}^{R}}},
\label{6}
\end{equation}
can replace an equation (\ref{3}) for
\begin{equation}
(i\gamma^{\mu}\partial_{\mu}^{s}-m_{s})\psi_{s}=0.
\label{7}
\end{equation}

In these circumstances the free Dirac Lagrangian of a C-noninvariant fermion becomes naturally 
united and behaves as 
\begin{equation}
L_{free}^{D}=\overline{\psi}_{s}\gamma^{5}(i\gamma^{\mu}\partial_{\mu}^{s}-m_{s})\psi_{s}.
\label{8}  
\end{equation} 

The latter does not imply of course that the mass, energy, and momentum of the neutrino of true neutrality at the level as were united by the author [3] in a unified whole do not transform 
the left (right)-handed axial-vector neutrino into a right (left)-handed one without violate 
of Lorentz symmetry. Thereby, it does not exclude [1] the fact that regardless of whether or not 
an unbroken Lorentz invariance exists, the same neutrino may not be simultaneously both a left-handed fermion and a right-handed one. This is exactly the same as when each type of gauge boson constitutes a kind of physical current. Consequently, the Lagrangian such as (\ref{8}) expresses, for each of the existing types of the local axial-vector gauge transformations, the idea of so far unobserved unified mirror principle.

Our purpose in a given work is to formulate this principle and its consequences by investigating 
the questions implied from the invariance of the free Dirac Lagrangian (\ref{8}) concerning the 
two forms of the local axial-vector gauge transformations including a unified theoretical description of the origination of mass of truly neutral types of particles and fields at the new level, namely, at the level of hitherto internally undisclosed structure of an axial-vector gauge invariance.

\vspace{0.8cm}
\noindent
{\bf 2. Mass structure of axial-vector types of photons}
\vspace{0.4cm}

The importance of our notion about an electric charge of a C-noninvariant nature lies in 
the fact [5] that between leptonic current structural components there exist some paradoxical contradictions, which admit their classification with respect to C-operation. This emphasizes 
the circumstance that the classical anapole [8,9] must be considered as the C-odd electric charge. 
It has a crucial value for axial-vector types of local gauge transformations.

One of them expresses, in the Coulomb $(C)$ limit, the idea about that
\begin{equation}
\psi'_{s}=U_{s}^{C}\psi_{s}, \, \, \, \, U_{s}^{C}=e^{i\beta_{s}(x_{s})\gamma^{5}},
\label{9}  
\end{equation} 
and the Lagrangian (\ref{8}) loses at the local phase $\beta_{s}(x_{s})$ his  
gauge invariance. 

For restoration of such a broken symmetry, it is desirable to introduce the photon Coulomb field
$A_{\mu}^{s}(x_{s})$ corresponding in a system to an axial-vector transformation
\begin{equation}
A_{\mu}^{s'}=A_{\mu}^{s}+\frac{i}{e_{s}}\gamma^{5}\partial_{\mu}^{s}\beta_{s}
\label{10}  
\end{equation} 
including the Coulomb mirror interaction constants $e_{s}$ at the level of an electric charge 
of the C-noninvariant nature.

Insertion of
\begin{equation}
\partial_{\mu}^{s}=\partial_{\mu}^{s}-e_{s}A_{\mu}^{s}
\label{11}  
\end{equation} 
in (\ref{8}) leads us to the Lagrangian 
$$L^{D}=L_{free}^{D}+L_{int}^{D}=$$
\begin{equation}
=\overline{\psi}_{s}\gamma^{5}(i\gamma^{\mu}\partial_{\mu}^{s}-m_{s})\psi_{s}-
ie_{s}\overline{\psi}_{s}\gamma^{5}\gamma^{\mu}\psi_{s}A_{\mu}^{s}.
\label{12}  
\end{equation}  

Its invariance concerning the acting local gauge transformations (\ref{9}) and (\ref{10}) becomes possible owing to the interaction with an axial-vector photon Coulomb field of truly neutral types 
of fermions. 

However, the fact that the very equation (\ref{7}) does not exclude the symmetry with respect 
to its matrix structure indicates the role of the unified principle in all Lagrangians of its structural objects. Therefore, if it turns out that one such an object may be any of $\gamma^{5},$ 
$\gamma^{\mu},$ $\partial_{\mu}^{s},$ $m_{s},$ $e_{s},$ $\overline{\psi}_{s},$ and $\psi_{s},$ the field $A_{\mu}^{s}$ equalized with the Coulomb field of an axial-vector photon $(\gamma^{A})$ must not in an interaction Lagrangian $L_{int}^{D}$ be usual two component field, because it can appear only in conformity with the acting quantum operator $\partial_{\mu}^{s}$ as a consequence of 
$4\times 4$ matrix, which is absent in a classical C-noninvariant Dirac Lagrangian.

So, we must recognize that the gauge state such as $A_{\mu}^{s}$ says about the existence in 
an axial-vector photon of a kind of inertial mass. In other words, each of C-odd left- or 
right-handed Coulomb field from 
\begin{equation}
A_{\mu}^{s}=\left(A_{\mu}\atop B_{\mu}\right), \, \, \, \,
A_{\mu}=\left(A_{\mu}^{L}\atop A_{\mu}^{R}\right), \, \, \, \,
B_{\mu}=\left(B_{\mu}^{L}\atop B_{\mu}^{R}\right)
\label{13}
\end{equation}
corresponds in a Lagrangian (\ref{12}) to a kind of fermion field [3] from 
\begin{equation}
\psi_{s}=\left(\psi\atop \phi\right), \, \, \, \,
\psi=\left(\psi_{L}\atop \psi_{R}\right), \, \, \, \,
\phi=\left(\phi_{L}\atop \phi_{R}\right).
\label{14}
\end{equation}

The coexistence, in the case of the Dirac Lagrangian (\ref{12}), of both types of fields (\ref{13}) and (\ref{14}) expresses the idea of the mentioned experiments [6,7] about neutrino scattering on a nucleus as an indication in favor of that the left (right)-handed axial-vector neutrinos due to the spontaneous mirror symmetry violation have no interaction with right (left)-handed photons of true neutrality. They possess with all the left (right)-handed C-odd gauge bosons the same interaction 
as the truly neutral electrons [1] of left (right) helicity.

It is already clear from the foregoing that the left- and right-handed axial-vector photons are 
of long- and short-lived bosons of true neutrality, respectively. Such a difference in lifetimes
of photons of definite helicity can explain the spontaneous mirror symmetry absence, which comes forward also in the universe of C-noninvariant types of particles and fields as a spontaneity criterion of gauge invariance violation [10]. Thereby, it says about axial-vector photons of the different components possessing the unidentical masses, energies, and momenta. These properties have important consequences for generalization of the classical Klein [11]-Gordon [12] equation from the quantum electrodynamics of spinless particles [13] to the case of C-odd particles with an integral spin, because they give the possibility to directly define the structure of the latter for the 
four-component wave function $\varphi_{s}(t_{s}, {\bf x}_{s})$ in a mirror world as following:
\begin{equation}
(\partial_{\mu}^{s}\partial^{\mu}_{s}+m_{s}^{2})\varphi_{s}=0
\label{15}
\end{equation}
in which appears one more connection such that
\begin{equation}
\varphi_{s}=\left(\varphi\atop \chi\right), \, \, \, \,
\varphi=\left(\varphi_{L}\atop \varphi_{R}\right), \, \, \, \,
\chi=\left(\chi_{L}\atop \chi_{R}\right).
\label{16}
\end{equation}

The Lagrangian responsible for an equation (\ref{15}) must have the form
\begin{equation}
L_{free}^{B}=
\frac{1}{2}\varphi_{s}^{*}\gamma^{5}(\partial_{\mu}^{s}\partial^{\mu}_{s}+m_{s}^{2})\varphi_{s}.
\label{17}  
\end{equation} 

In this definition, we have used the matrix $\gamma^{5},$ because a set-matrix duality principle [14] is the unified for all the quantum theory equations. Insofar as its role allowing us to reveal the structural connections of the mirror equation (\ref{15}) and to include in the discussion their aspects is concerned, it calls for special presentation.

At a choice of an axial-vector gauge transformation 
\begin{equation}
\varphi'_{s}=U_{s}^{C}\varphi_{s}, \, \, \, \, U_{s}^{C}=e^{i\beta_{s}(x_{s})\gamma^{5}}
\label{18}  
\end{equation} 
having his own local phase $\beta_{s}(x_{s}),$ the invariance of the free boson Lagrangian 
(\ref{17}) is violated without loss of possibility of further restoration. We can, therefore,  introduce an axial-vector field $A_{\mu}^{s}$ in the presence of a kind of gauge transformation.

Unification of (\ref{11}) and (\ref{17}) convinces us here that at the local axial-vector  
transformations (\ref{10}) and (\ref{18}), the Lagrangian
$$L^{B}=L_{free}^{B}+L_{int}^{B}=$$
$$=\frac{1}{2}\varphi_{s}^{*}\gamma^{5}(\partial_{\mu}^{s}\partial^{\mu}_{s}+m_{s}^{2})\varphi_{s}+$$
\begin{equation}
+\frac{1}{2}[e_{s}(\varphi_{s}^{*}\gamma^{5}\partial_{\mu}^{s}\varphi_{s}A^{\mu}_{s}-
\varphi_{s}^{*}\gamma^{5}\partial^{\mu}_{s}\varphi_{s}A_{\mu}^{s})-
e_{s}^{2}\varphi_{s}^{*}\gamma^{5}\varphi_{s}A_{\mu}^{s}A^{\mu}_{s}]
\label{19}  
\end{equation} 
steel remains gauge-invariant. 

\vspace{0.8cm}
\noindent
{\bf 3. Axial-vector photon fields of Coulomb and Newton nature}
\vspace{0.4cm}

The Lagrangian $L_{int}^{D}$ as a part of the Dirac axial-vector interaction corresponds in (\ref{12}) to the C-noninvariant electric charges of the interacting objects. But, as was noted 
in [15] for the first time, any interaction between the fermion and the field of emission includes not only a kind of Coulomb part but also a kind of Newton part. This in turn implies the coexistence, in the case of the Lagrangian (\ref{12}), of both types of the Dirac interaction structural components. Such a connection is realized owing to a mass-charge duality [16], according to which, each of the electric $E,$ weak $W,$ strong $S,$ and other innate types of charges testifies in favor of a kind of inertial mass. The masses and charges of a C-odd particle constitute herewith the united rest mass $m_{s}^{U}$ and charge $e_{s}^{U}$ coinciding with all its mass and charge:
\begin{equation}
m_{s}=m_{s}^{U}=m_{s}^{E}+m_{s}^{W}+m_{s}^{S}+...,
\label{20}
\end{equation}
\begin{equation}
e_{s}=e_{s}^{U}=e_{s}^{E}+e_{s}^{W}+e_{s}^{S}+....
\label{21}
\end{equation}

We encounter, thus, the fact that nature itself characterizes each free Lagrangian both from the point of view of charge and from the point of view of mass. Thereby, it admits the existence in any free Dirac or boson Lagrangian not only of a kind of Coulomb $(C)$ component but also of a kind of Newton $(N)$ component, the invariance of each of which concerning a kind of local gauge transformation leads to the appearance of the same interaction corresponding part. 

Tnis correspondence principle may serve, in the limits of $m_{s}=m_{s}^{E}$ and $e_{s}=e_{s}^{E},$ 
as an invariance criterion of any Lagrangian from (\ref{12}) and (\ref{21}) concerning the action of one more another type of the local axial-vector gauge transformation. One can define the structure 
of such a second type of an axial-vector transformation, which has the different local phase 
$\beta_{s}(\tau_{s})$ for fermion $\psi_{s}$ and boson $\varphi_{s}$ fields by the following manner:
\begin{equation}
\psi'_{s}=U_{s}^{N}\psi_{s}, \, \, \, \, U_{s}^{N}=e^{i\beta_{s}(\tau_{s})\gamma^{5}},
\label{22}  
\end{equation} 
\begin{equation}
\varphi'_{s}=U_{s}^{N}\varphi_{s}, \, \, \, \, U_{s}^{N}=e^{i\beta_{s}(\tau_{s})\gamma^{5}}.
\label{23}  
\end{equation} 

One more characteristic moment is that the Newton component with mass $m_{s}$ in any Lagrangian from (\ref{8}) and (\ref{17}) is general concerning the corresponding axial-vector gauge transformation (\ref{9}) or (\ref{18}) and does not depend of whether it has a local or a global phase.
Additionally, all conditions of gauge symmetry of a Coulomb part with operator $\partial_{\mu}^{s}$ hold in each Lagrangian from (\ref{8}) and (\ref{17}) regardless of whether the suggested second type 
of an axial-vector transformation is or not present in it.

Therefore, to conform with all quantum operators (\ref{5}) and (\ref{6}), we must choose a particle mass $m_{s}=-i\partial_{\tau}^{s}$ in which 
\begin{equation}
\partial_{\tau}^{s}=
{{0 \ \, \, \, \, \partial_{\tau}^{A}}\choose{\partial_{\tau}^{A} \, \, \, \, \ 0}}, 
\, \, \, \, \partial_{\tau}^{A}=
{{\partial_{\tau}^{L} \, \, \, \, 0}\choose{\ 0 \, \, \, \, \ \partial_{\tau}^{R}}}.
\label{24}
\end{equation}

From their point of view, the free Dirac Lagrangian (\ref{8}) accepts the naturally united form
\begin{equation}
L_{free}^{D}=
i\overline{\psi}_{s}\gamma^{5}(\gamma^{\mu}\partial_{\mu}^{s}+\partial_{\tau}^{s})\psi_{s}
\label{25}  
\end{equation} 
responsible for an equation 
\begin{equation}
(\gamma^{\mu}\partial_{\mu}^{s}+\partial_{\tau}^{s})\psi_{s}=0.
\label{26}
\end{equation}

At first sight, the latter says that either $\psi_{s}$ comes forward in it as a function 
$\psi_{s}(t_{s}, {\bf x}_{s}, \tau_{s}),$ namely, as a function of the fermion space-time 
coordinates $(t_{s}, {\bf x}_{s})$ and lifetimes $\tau_{s}$ or our reasoning about the quantum operator presentation of mass is not valid. Such an implication, however, does not correspond to reality. The point is that the spin properties of a truly neutral particle depend not only on its axial-vector mass, energy, and momentum but also on the nature of space [3], which characterizes it by space-time coordinates and lifetime until this is forbidden by symmetry laws. Therefore, without loss of generality, the axial-vector operators $\partial_{\mu}^{s}$ and $\partial_{\tau}^{s}$ can individually influence on the fermion field $\psi_{s}$ as well as on each of the existing types of 
wave functions. By the same reason, we conclude that 
\begin{equation}
\partial_{\mu}^{s}\psi_{s}=\partial_{\mu}^{s}\psi_{s}(x_{s}), \, \, \, \, 
\partial_{\tau}^{s}\psi_{s}=\partial_{\tau}^{s}\psi_{s}(\tau_{s}),
\label{27}
\end{equation}
\begin{equation}
\partial_{\mu}^{s}\beta_{s}=\partial_{\mu}^{s}\beta_{s}(x_{s}), \, \, \, \, 
\partial_{\tau}^{s}\beta_{s}=\partial_{\tau}^{s}\beta_{s}(\tau_{s}).
\label{28}
\end{equation}

With the use of the second type of an axial-vector transformation (\ref{22}), the Newton 
part with operator $\partial_{\tau}^{s}$ must lead to the appearance in a Lagrangian (\ref{25}) 
of one of its gauge-noninvariant components and that, consequently, the further restoration of such 
a broken symmetry requires one to introduce the Newton field $A_{\tau}^{s}(\tau_{s})$ in conformity with an axial-vector transformation
\begin{equation}
A_{\tau}^{s'}=A_{\tau}^{s}+\frac{i}{m_{s}}\gamma^{5}\partial_{\tau}^{s}\beta_{s}
\label{29}  
\end{equation} 
including the Newton mirror interaction constants $m_{s}$ at the level of an electric 
mass of the axial-vector nature.

Taking into account (\ref{11}), (\ref{27}), (\ref{28}) and that
\begin{equation}
\partial_{\tau}^{s}=\partial_{\tau}^{s}-m_{s}A_{\tau}^{s},
\label{30}  
\end{equation} 
from (\ref{25}), we are led to another new Lagrangian $L^{D},$ which is invariant concerning 
the local axial-vector gauge transformations (\ref{9}), (\ref{10}), (\ref{22}), and (\ref{29}), because it consists of the C-odd Dirac interaction following parts:
$$L^{D}=L_{free}^{D}+L_{int}^{D}=$$
\begin{equation}
=i\overline{\psi}_{s}\gamma^{5}(\gamma^{\mu}\partial_{\mu}^{s}+\partial_{\tau}^{s})\psi_{s}-
ie_{s}j^{\mu}_{C}A_{\mu}^{s}-im_{s}j^{\tau}_{N}A_{\tau}^{s}.
\label{31}  
\end{equation} 
Here $j^{\mu}_{C}$ and $j^{\tau}_{N}$ describe the Coulomb and Newton components of the same
axial-vector photon leptonic current
\begin{equation}
j^{\mu}_{C}=\overline{\psi}_{s}\gamma^{5}\gamma^{\mu}\psi_{s},
\label{32}  
\end{equation} 
\begin{equation}
j^{\tau}_{N}=\overline{\psi}_{s}\gamma^{5}\psi_{s}.
\label{33}  
\end{equation} 

One of the most highlighted features of these types of currents is their unity, which 
involves the commutativity conditions of matrices $\gamma^{5},$ $\partial_{\mu}^{s},$ and 
$\partial_{\tau}^{s}$ expressing the idea of the coexistence law of the continuity equations
\begin{equation}
\partial_{\mu}^{s}j^{\mu}_{C}=0,
\label{34}  
\end{equation} 
\begin{equation}
\partial_{\tau}^{s}j^{\tau}_{N}=0.
\label{35}  
\end{equation} 

Simultaneously, as is easy to see, the field $A_{\mu}^{s}$ and 
\begin{equation}
A_{\tau}^{s}=\left(A_{\tau}\atop B_{\tau}\right), \, \, \, \,
A_{\tau}=\left(A_{\tau}^{L}\atop A_{\tau}^{R}\right), \, \, \, \,
B_{\tau}=\left(B_{\tau}^{L}\atop B_{\tau}^{R}\right)
\label{36}
\end{equation}
arise in a Lagrangian (\ref{31}) as the Coulomb and Newton parts of the same 
axial-vector photon field. 

The structure itseif of the Lagrangian $L_{int}^{D}$ in (\ref{31}) testifies in addition that at the availability of an interaction Newton component with an axial-vector photon field, the neutrino of true neutrality must possess an axial-vector electric mass. Insofar as its C-noninvariant electric charge is concerned, it appears in the Coulomb part dependence of the same interaction.

The quantum mass operator in turn transforms (\ref{17}) into a latent united Lagrangian of 
the unified field theory of C-odd bosons with a nonzero spin
\begin{equation}
L_{free}^{B}=\frac{1}{2}\varphi_{s}^{*}\gamma^{5}
(\partial_{\mu}^{s}\partial^{\mu}_{s}-\partial_{\tau}^{s}\partial^{\tau}_{s})\varphi_{s}.
\label{37}  
\end{equation} 

This connection suggests another new equation 
\begin{equation}
(\partial_{\mu}^{s}\partial^{\mu}_{s}-\partial_{\tau}^{s}\partial^{\tau}_{s})\varphi_{s}=0
\label{38}
\end{equation}
and thereby confirms the fact that the operators $\partial_{\mu}^{s}$ and $\partial_{\tau}^{s}$ 
can individually influence not only on the fermion but also on the boson $\varphi_{s}$ wave function 
\begin{equation}
\partial_{\mu}^{s}\varphi_{s}=\partial_{\mu}^{s}\varphi_{s}(x_{s}), \, \, \, \, 
\partial_{\tau}^{s}\varphi_{s}=\partial_{\tau}^{s}\varphi_{s}(\tau_{s}).
\label{39}
\end{equation}  

Uniting (\ref{37}) with (\ref{11}), (\ref{30}) and having in mind (\ref{28}) and (\ref{39}), we 
find that the Lagrangian $L^{B}$ invariant concerning the local axial-vector gauge transformations 
(\ref{9}), (\ref{10}), (\ref{23}), and (\ref{29}) must contain the following components of 
the C-noninvariant boson interaction:
$$L^{B}=L_{free}^{B}+L_{int}^{B}=$$
$$=\frac{1}{2}\varphi_{s}^{*}\gamma^{5}
(\partial_{\mu}^{s}\partial^{\mu}_{s}-\partial_{\tau}^{s}\partial^{\tau}_{s})\varphi_{s}+$$
$$+\frac{1}{2}[e_{s}(J_{\mu}^{C}A^{\mu}_{s}-J^{\mu}_{C}A_{\mu}^{s})-
e_{s}^{2}\varphi_{s}^{*}\gamma^{5}\varphi_{s}A_{\mu}^{s}A^{\mu}_{s}]-$$
\begin{equation}
-\frac{1}{2}[m_{s}(J_{\tau}^{N}A^{\tau}_{s}-J^{\tau}_{N}A_{\tau}^{s})-
m_{s}^{2}\varphi_{s}^{*}\gamma^{5}\varphi_{s}A_{\tau}^{s}A^{\tau}_{s}].
\label{40}  
\end{equation} 

Coulomb and Newton electric currents, $J^{\mu}_{C}(J_{\mu}^{C})$ and $J^{\tau}_{N}(J_{\tau}^{N}),$ interacting with axial-vector fields $A_{\mu}^{s}(A^{\mu}_{s})$ and $A_{\tau}^{s}(A^{\tau}_{s})$
respectively, constitute the two parts of the same axial-vector boson current
\begin{equation}
J^{\mu}_{C}=\varphi_{s}^{*}\gamma^{5}\partial^{\mu}_{s}\varphi_{s}, \, \, \, \,
J_{\mu}^{C}=\varphi_{s}^{*}\gamma^{5}\partial_{\mu}^{s}\varphi_{s},
\label{41}  
\end{equation} 
\begin{equation}
J^{\tau}_{N}=\varphi_{s}^{*}\gamma^{5}\partial^{\tau}_{s}\varphi_{s}, \, \, \, \,
J_{\tau}^{N}=\varphi_{s}^{*}\gamma^{5}\partial_{\tau}^{s}\varphi_{s}.
\label{42}  
\end{equation} 

As well as in the systems of axial-vector leptonic currents, conservation of each boson current
here must carry out itself as a consequence of the coexistence law of the continuity equations
\begin{equation}
\partial_{\mu}^{s}J^{\mu}_{C}=0, \, \, \, \, \partial^{\mu}_{s}J_{\mu}^{C}=0,
\label{43}  
\end{equation} 
\begin{equation}
\partial_{\tau}^{s}J^{\tau}_{N}=0, \, \, \, \, \partial^{\tau}_{s}J_{\tau}^{N}=0.
\label{44}  
\end{equation} 

In particular, about sizes of $m_{s}$ and $e_{s}$ should be mentioned, corresponding in a 
Lagrangian (\ref{40}) to the fact that owing to the interaction with Newton $A_{\tau}^{s}$ 
and Coulomb $A_{\mu}^{s}$ fields of the same axial-vector photon, any of truly neutral bosons 
with a nonzero spin must possess simultaneously each of the axial-vector types of electric mass 
and charge. Such a boson can, for example, be photon itself. It is not surprising therefore that $m_{s}^{2}A_{\tau}^{s}A^{\tau}_{s}$ and $e_{s}^{2}A_{\mu}^{s}A^{\mu}_{s}$ describe in (\ref{40}) 
its Newton and Coulomb interactions with another photon of an axial-vector nature.

In both Lagrangians (\ref{31}) and (\ref{40}), as is now well known, the mass $m_{s}$ and charge $e_{s},$ each of which is present jointly with a kind of axial-vector photon field, appear in the mass-charge structure dependence of gauge invariance, so that there exist the axial-vector tensors
\begin{equation}
F_{\mu\lambda}^{C}=\partial_{\mu}A_{\lambda}^{C}-\partial_{\lambda}A_{\mu}^{C},
\label{45}
\end{equation}
\begin{equation}
F_{\tau\sigma}^{N}=\partial_{\tau}A_{\sigma}^{N}-\partial_{\sigma}A_{\tau}^{N}.
\label{46}
\end{equation}

With their availability, the structure of the unified field theory Lagrangian of C-odd neutrinos and particles with an integral spin has fully definite form
$$L=
i\overline{\psi}_{s}\gamma^{5}(\gamma^{\mu}\partial_{\mu}^{s}+\partial_{\tau}^{s})\psi_{s}+$$
$$+\frac{1}{2}\varphi_{s}^{*}\gamma^{5}
(\partial_{\mu}^{s}\partial^{\mu}_{s}-\partial_{\tau}^{s}\partial^{\tau}_{s})\varphi_{s}-$$
$$-\frac{1}{4}F_{\mu\lambda}^{C}F^{\mu\lambda}_{C}+
\frac{1}{4}F_{\tau\sigma}^{N}F^{\tau\sigma}_{N}-
ie_{s}j^{\mu}_{C}A_{\mu}^{s}-im_{s}j^{\tau}_{N}A_{\tau}^{s}+$$
$$+\frac{1}{2}[e_{s}(J_{\mu}^{C}A^{\mu}_{s}-J^{\mu}_{C}A_{\mu}^{s})-
e_{s}^{2}\varphi_{s}^{*}\gamma^{5}\varphi_{s}A_{\mu}^{s}A^{\mu}_{s}]-$$
\begin{equation}
-\frac{1}{2}[m_{s}(J_{\tau}^{N}A^{\tau}_{s}-J^{\tau}_{N}A_{\tau}^{s})-
m_{s}^{2}\varphi_{s}^{*}\gamma^{5}\varphi_{s}A_{\tau}^{s}A^{\tau}_{s}].
\label{47}  
\end{equation} 

Its components reflect just the fact that each of the axial-vector types of fermions or bosons 
possesses the C-noninvariant electric charge at the interaction with the Coulomb $A_{\mu}^{s}$ field 
of an axial-vector photon. This in turn implies the origination of an axial-vector electric mass of the investigated particles as a consequence of their interaction with the Newton $A_{\tau}^{s}$
field of the same type of photon.

We recognize that a characteristic part of the standard model [17-19] is the chiral presentation 
of the Weyl [20] in which a matrix $\gamma_{5}$ allows one to choose only the left components of 
the fermion field. In this situation, the presence of mass of any particle in an interaction Lagrangian violates its gauge invariance.

At first sight, such a violation requires the existence of one more another type of the scalar 
boson [21] responsible for origination in a Lagrangian of mass of the interacting particles and fields. This, however, is not in line with nature, since the availability in it of the second type 
of the local transformation expressing the idea of the mass structure [22] of gauge invariance
has not been known before the creation of the first-initial electroweak theory.

\vspace{0.8cm}
\noindent
{\bf 4. Conclusion}
\vspace{0.4cm}

Another of the most highlighted features of axial-vector mass, energy, and momentum is such 
that they together with a relation
\begin{equation}
E_{s}=\frac{{\bf p}_{s}^{2}}{2m_{s}}
\label{48}
\end{equation}
constitute the C-noninvariant Schr\"odinger equation 
\begin{equation}
i\frac{\partial \psi_{s}}{\partial t_{s}}-
\frac{1}{2m_{s}\psi_{s}}\frac{\partial\psi_{s}}{\partial {\bf x}_{s}}
\frac{\partial\psi_{s}}{\partial {\bf x}_{s}}=0.
\label{49}
\end{equation}

With the availability of a particle quantum mass operator $m_{s}=-i\partial_{\tau}^{s},$ it
suggests one more highly important equation
\begin{equation}
\partial_{t}^{s}\psi_{s}\partial_{\tau}^{s}\psi_{s}-
\frac{1}{2}\partial_{\bf x}^{s}\psi_{s}\partial_{\bf x}^{s}\psi_{s}=0.
\label{50}
\end{equation}

As well as in (\ref{26}) and (\ref{38}), the axial-vector operators $\partial_{t}^{s}$ and 
$\partial_{\bf x}^{s}$ can individually influence here on the matter field
\begin{equation}
\partial_{t}^{s}\psi_{s}=\partial_{t}^{s}\psi_{s}(t_{s}), \, \, \, \,
\partial_{\bf x}^{s}\psi_{s}=\partial_{\bf x}^{s}\psi_{s}({\bf x}_{s}).
\label{51}
\end{equation}

Turning again to the equation (\ref{7}), we remark that to it one can also lead by another way inserting the Lagrangian (\ref{8}) in the Euler-Lagrange [23] equation, which in a mirror world 
has the fully regular form
\begin{equation}
\partial_{\mu}^{s}\left(\frac{\partial L_{free}^{D}}
{\partial(\partial_{\mu}^{s}\overline{\psi}_{s})}\right)=
\frac{\partial L_{free}^{D}}{\partial\overline{\psi}_{s}}.
\label{52}
\end{equation}

Furthermore, if it turns out that the latter is not in state to establish (\ref{26}) on account 
of (\ref{25}) and (\ref{27}), this says about the existence in the same action of both Coulomb and Newton parts [15]. They constitute in a mirror presentation the Euler-Lagrange equation at the new level, namely, at the level of the mass-charge structure of an axial-vector gauge invariance:
\begin{equation}
\partial_{\mu}^{s}\left(\frac{\partial L_{free}^{D}}
{\partial(\partial_{\mu}^{s}\overline{\psi}_{s}(x_{s}))}\right)+
\partial_{\tau}^{s}\left(\frac{\partial L_{free}^{D}}
{\partial(\partial_{\tau}^{s}\overline{\psi}_{s}(\tau_{s}))}\right)=
\frac{\partial L_{free}^{D}}{\partial\overline{\psi}_{s}(x_{s})}+
\frac{\partial L_{free}^{D}}{\partial\overline{\psi}_{s}(\tau_{s})}.
\label{53}
\end{equation}

It states that 
\begin{equation}
\partial_{\tau}^{s}\partial_{\mu}^{s}\psi_{s}(x_{s})=0, \, \, \, \, 
\partial_{\tau}^{s}\psi_{s}(\tau_{s})\partial_{\mu}^{s}\psi_{s}(x_{s})\neq 0,
\label{54}
\end{equation}
\begin{equation}
\partial_{\mu}^{s}\partial_{\tau}^{s}\psi_{s}(\tau_{s})=0, \, \, \, \,
\partial_{\mu}^{s}\psi_{s}(x_{s})\partial_{\tau}^{s}\psi_{s}(\tau_{s})\neq 0.
\label{55}
\end{equation}

A fundamental role in nature of nonelectric components of the axial-vector mass and charge of truly neutral neutrinos and bosons with a nonzero spin and some above unnoted aspects of new equations of their unified field theory call for special presentation. 

\vspace{0.8cm}
\noindent
{\bf References}
\begin{enumerate}
\item
R.S. Sharafiddinov, Int. J. Theor. Phys. {\bf 55}, 3040 (2016), doi: 10.1007/s10773-016-2936-8. 
Bull. Am. Phys. Soc. {\bf 57}(16), KA.00069 (2012); e-print arXiv:1004.0997 [hep-ph].
\item
R.S. Sharafiddinov, Bull. Am. Phys. Soc. 59(5), L1.00036 (2014).
\item
R.S. Sharafiddinov, Int. J. Theor. Phys. {\bf 55}, 2139 (2016), doi: 10.1007/s10773-015-2852-3.
Bull. Am. Phys. Soc. {\bf 61}(5), T1.00035 (2016); e-print arXiv:1506.01040 [physics.gen-ph].
\item
R.S. Sharafiddinov, Can. J. Phys. {\bf 93}, 1005 (2015), doi: 10.1139/cjp-2014-0497. 
Bull. Am. Phys. Soc. {\bf 59}(18), EC.00001 (2014); e-print arXiv:1409.2397 [physics.gen-ph].
\item
R.S. Sharafiddinov, J. Phys. Nat. Sci. {\bf 4}, 1 (2013); physics/0702233.
\item
K.S. Kuzmin, V.V. Lyubushkin and V.A. Naumov, Eur. Phys. J. {\bf C 54}, 517 (2008); 
0712.4384 [hep-ph].
\item
The CMS Collaboration, Phys. Rev. {\bf D 74}, 052002 (2006); hep-ex/0603034. 
\item
Ya.B. Zel'dovich, J. Exp. Theor. Phys. {\bf 33}, 1531 (1957).
\item
Ya.B. Zel'dovich and A.M. Perelomov, J. Exp. Theor. Phys. {\bf 39}, 1115 (1960).
\item
A. Giveon and E. Witten, Phys. Lett. {\bf B 332}, 44 (1994); 
e-print arXiv:hep-th/9404184.
\item
O. Klein, Z. Phys. {\bf 37}, 895 (1926).
\item 
W. Gordon, Z. Phys. {\bf 40}, 117 (1926-1927).
\item
W. Pauli and V. Weisskopf, Helv. Phys. Acta, {\bf 7}, 709 (1934). 
\item
R.S. Sharafiddinov, Bull. Am. Phys. Soc. 63(4), K14.00008 (2018); 
e-print arXiv:1612.01828 [physics.gen-ph].
\item
R.S. Sharafiddinov, Spacetime Subst. {\bf 2}, 87 (2003). Bull. Am. Phys. Soc. {\bf 59}(18), 
EC.00006 (2014); e-print arXiv:hep-ph/0401230.
\item
R.S. Sharafiddinov, Spacetime Subst. {\bf 3}, 47 (2002). Bull. Am. Phys. Soc. 59(5), 
T1.00009 (2014); e-print arXiv:physics/0305008.
\item
S.L. Glashow, Nucl. Phys. {\bf 22}, 579 (1961).
\item
A. Salam and J.C. Ward, Phys. Lett. {\bf 13}, 168 (1964).
\item
S. Weinberg, Phys. Rev. Lett. {\bf 19}, 1264 (1967).
\item
H. Weyl, Z. Phys. {\bf 56,} 330 (1929), doi: 10.1007/BF01339504. 
\item
P.W. Higgs, Phys. Rev. Lett. {\bf 13}, 508 (1964).
\item
R.S. Sharafiddinov, Phys. Essays {\bf 19}, 58 (2006); e-print arXiv:hep-ph/0407262.
\item
V.S. Vladimirov, et al., Collection of Problems on Mathematical Physics Equations 
[in Russian], Nauka, Moscow, 1974.
\end{enumerate}
\end{document}